\documentclass[sigconf,nonacm]{acmart}

\renewcommand\footnotetextcopyrightpermission[1]{}
\usepackage{algorithm}
\usepackage{algorithmic}
\usepackage{multirow}

\AtBeginDocument{%
  }

\setcopyright{none}

\begin{document}

\title{SDSL-Solver: Scalable Distributed Sparse Linear Solvers for Large-Scale Interior Point Methods}


\author{Shaofeng Yang}
\affiliation{%
  \institution{Institute of Computing Technology, Chinese Academy of Science}
  \city{Beijing}
  \country{China}}
\email{yangshaofeng@ncic.ac.cn}

\author{Yunting Wang}
\affiliation{%
  \institution{Institute of Computing Technology, Chinese Academy of Science}
  \city{Beijing}
  \country{China}}
\email{wangyunting@ncic.ac.cn}

\author{Yingying Cheng}
\affiliation{%
 \institution{Theory Lab, \\ Huawei Technologies Co., Ltd.}
 \city{Beijing}
 \country{China}}
\email{cheng.yingying1@huawei.com}

\author{Fan Zhang}
\affiliation{%
 \institution{Theory Lab, \\ Huawei Technologies Co., Ltd.}
 \city{Hong Kong}
 \country{China}}
\email{zhang.fan2@huawei.com}

\author{Xin He}
\affiliation{%
  \institution{Institute of Computing Technology, Chinese Academy of Science}
  \city{Beijing}
  \country{China}}
\email{hexin2016@ncic.ac.cn}

\author{Guangming Tan}
\affiliation{%
  \institution{Institute of Computing Technology, Chinese Academy of Science}
  \city{Beijing}
  \country{China}}
\email{tgm@ncic.ac.cn}




\begin{abstract}
The solution of sparse linear systems constitutes the dominant computational bottleneck in interior point methods (IPMs), frequently consuming over 70\% of the total solution time. As optimization problems scale to millions of variables, direct solvers encounter prohibitive fill-in, excessive memory consumption, and limited parallel scalability. We present SDSL-Solver, a scalable distributed sparse linear solver framework designed for IPMs. SDSL-Solver employs Krylov subspace methods, combined with numerics-based sparse filtering and diagonal correction techniques that produce high-quality preconditioners. To accommodate diverse problem characteristics, SDSL-Solver offers two complementary distributed parallel methods: Block Jacobi for diagonally dominant matrices, and Bordered Block Diagonal (BBD) for general or ill-conditioned matrices requiring globally coupled preconditioning via Schur complement techniques. A preconditioner reuse strategy further amortizes construction costs across consecutive IPMs iterations. We evaluate SDSL-Solver on benchmark problems with matrix dimensions ranging from tens of thousands to over five million on multi-node clusters equipped with X86 processors. The experimental results show that under the Block Jacobi and BBD distributed methods, SDSL-Solver on a four-node configuration achieves average speedups of $6.23\times$ and $7.77\times$, respectively, compared to PETSc running on the same number of nodes. Relative to the single-node PARDISO, the average speedups reach $97.54\times$ and $5.85\times$, respectively.

\end{abstract}




\maketitle

\section{Introduction}

\begin{figure}[t]
  \centering
  \includegraphics[width=0.75\columnwidth]{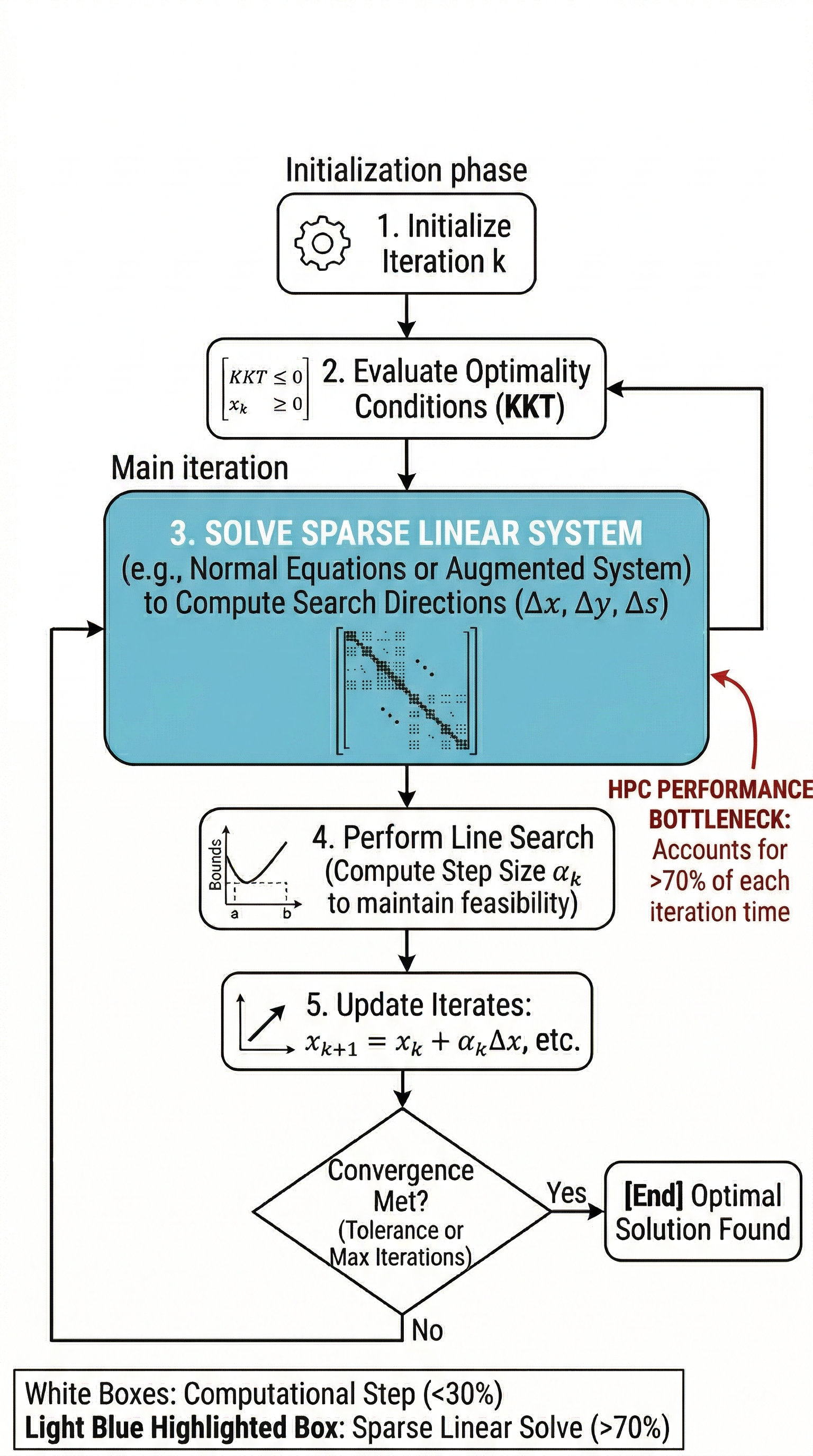}
  \caption{Workflow of an interior point method. The sparse linear system solve (highlighted in blue) dominates the total computation time, accounting for over 70\% of each iteration.}
  \label{fig:ipm_flow}
\end{figure}

Interior point methods (IPMs) constitute one of the most effective algorithmic frameworks for solving large-scale linear programming (LP), quadratic programming, and general convex optimization problems~\cite{wright1997primal,nocedal2006numerical}. At each iteration, the dominant computational cost resides in solving a sequence of sparse linear systems of the form $Ax = b$, which typically accounts for over 70\% of the total solution time~\cite{gondzio2012interior}. As optimization problems arising from real-world applications, including network flow, logistics, machine learning, and engineering design, scale to millions of variables and constraints, the efficient and robust solution of these linear systems becomes the critical bottleneck governing the overall scalability of IPMs, as illustrated in Figure~\ref{fig:ipm_flow}.

In the standard IPMs formulation, the coefficient matrix $A$ exhibits a structured block form amenable to reduction via Schur complement techniques, yielding a smaller system involving the matrix $S$. While this reduction is effective in many cases, the presence of dense columns in the constraint matrix can render the Schur complement $S$ nearly dense, thereby negating the computational benefits of the reduction. Moreover, as the IPMs progresses toward optimality, the diagonal scaling matrices introduce increasingly extreme values, causing the condition number of $S$ to grow by several orders of magnitude. This progressive ill-conditioning poses severe challenges for both the accuracy and convergence of the underlying linear solver.

Direct methods, exemplified by LU and Cholesky factorizations~\cite{davis2006direct}, represent the traditional approach for solving these systems. Although they provide broad applicability and strong numerical stability, direct methods incur significant fill-in during the factorization of sparse matrices, resulting in prohibitive memory consumption and computational cost at large scale. Furthermore, the inherently sequential nature of triangular solves constrains the parallel scalability of direct methods on modern multi-node architectures. For matrices with millions of rows, the fill-in can exceed the representable range of standard integer indexing, rendering direct solvers entirely inapplicable.

Krylov subspace iterative methods~\cite{saad2003iterative}, such as the Conjugate Gradient (CG)~\cite{cghestenes1952methods}, GMRES~\cite{gmressaad1986gmres}, BiCGSTAB~\cite{bicgstabvan1992bi}, and the Generalized Conjugate Residual (GCR)~\cite{gcreisenstat1983variational} method, provide a compelling alternative for large-scale sparse systems. These methods construct approximate solutions within expanding Krylov subspaces and rely primarily on sparse matrix-vector products and inner products, operations that are highly amenable to parallelization. However, for the ill-conditioned systems encountered in IPMs, unpreconditioned Krylov methods often exhibit unacceptably slow convergence or may fail to converge entirely. Effective preconditioning is therefore indispensable, yet constructing preconditioners that simultaneously offer high approximation quality, low application cost, and good parallel scalability remains an open challenge.

Existing parallel sparse solvers, such as PETSc~\cite{balay2019petsc} with Block Jacobi preconditioning~\cite{saad2003iterative}, provide a natural distributed framework by partitioning the matrix across processors and applying local preconditioners independently. Although conceptually straightforward and scalable, this approach neglects the coupling between off-diagonal blocks, leading to degraded convergence as the number of processors increases, a critical limitation for difficult, ill-conditioned problems. Conversely, direct parallel solvers such as PARDISO~\cite{schenk2004solving} deliver high accuracy but encounter the aforementioned scalability barriers for truly large-scale systems.

In this paper, we present SDSL-Solver, a comprehensive distributed parallel framework for solving the sparse linear systems arising in IPMs. Figure~\ref{fig:xsolver_arch} illustrates the overall architecture of SDSL-Solver, which unifies an advanced Krylov solver, adaptive preconditioning techniques, and two complementary distributed parallel methods within a single framework. Our approach addresses the limitations of existing methods through the following synergistic contributions:

\begin{figure*}[t]
  \centering
  \includegraphics[width=0.95\textwidth]{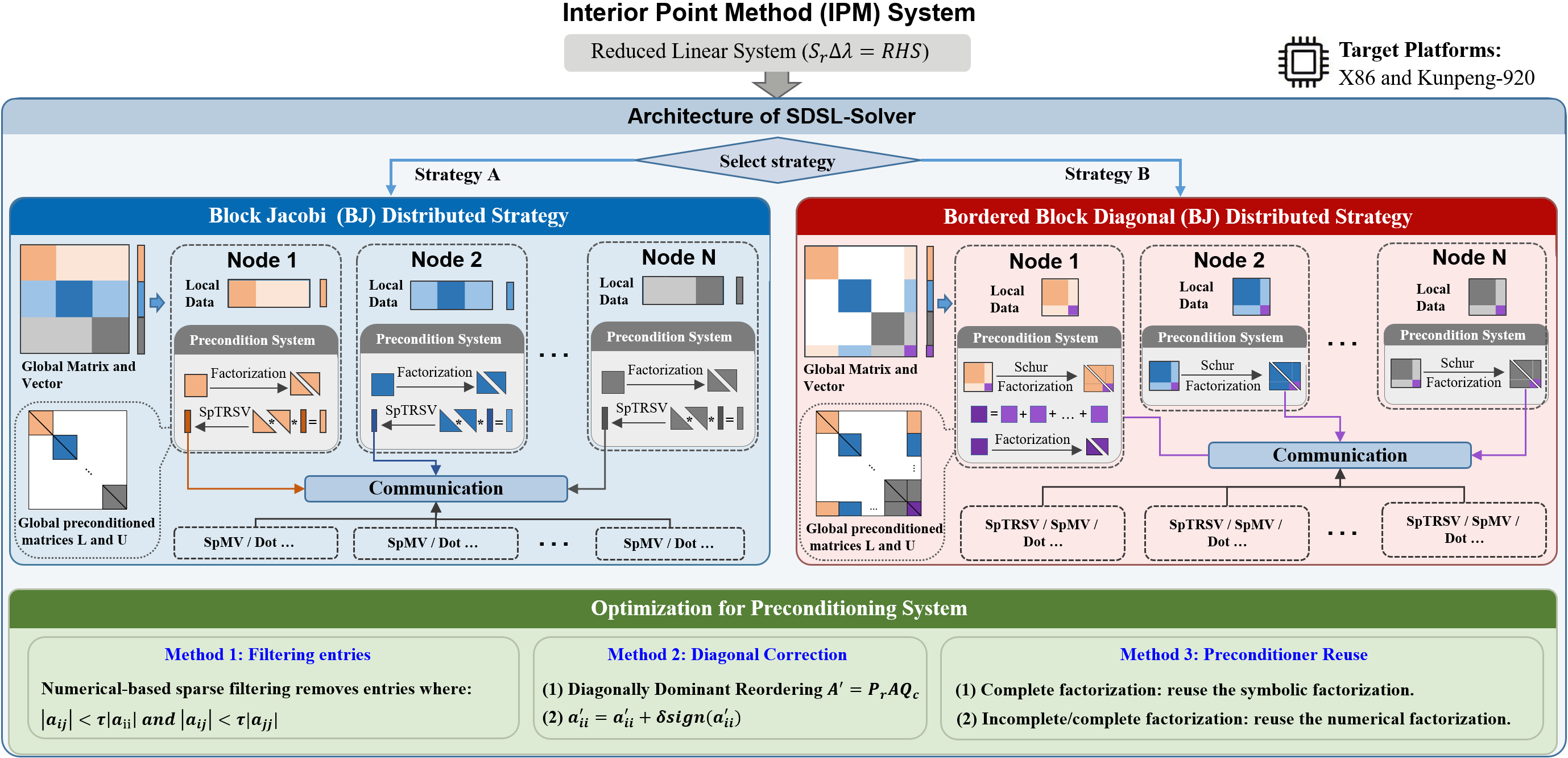}
  \caption{Architecture of SDSL-Solver. The framework receives the reduced linear system from the IPMs system and selects between Block Jacobi (for well-conditioned problems) and BBD (for ill-conditioned problems) distributed methods, both built on adaptive preconditioning with sparse filtering, diagonal correction, and preconditioner reuse.}
  \label{fig:xsolver_arch}
\end{figure*}

\begin{itemize}
  \item We design and implement two complementary distributed parallel methods: a \emph{Block Jacobi} method for diagonally dominant matrices that offers straightforward scalability, and a \emph{Bordered Block Diagonal (BBD)} method for general or ill-conditioned matrices that preserves global coupling information through a shared interface system solved via Schur complement techniques. A dynamic switching strategy automatically transitions between the two methods based on runtime solver diagnostics.

   \item We propose a \emph{numerics-based sparse filtering} algorithm that constructs high-quality preconditioners by selectively dropping small off-diagonal entries relative to the diagonal.
   
   \item We employ an MC64~\cite{2001On,hsl_mc64} algorithm to maximize the diagonal elements and a \emph{diagonal correction} technique that strengthens diagonal dominance to further improve conditioning. Together, these methods enable the use of incomplete factorization preconditioners (ILU/IC) while maintaining sufficient accuracy for IPMs convergence.

  \item We further introduce a \emph{preconditioner reuse} strategy that exploits the slow variation of coefficient matrices across consecutive IPMs iterations, amortizing the cost of preconditioner construction over multiple solves.
\end{itemize}

We conduct an extensive evaluation on a diverse set of benchmark problems with matrix dimensions ranging from tens of thousands to over five million. 
Experiments on X86 multi-node clusters show that on a four-node configuration, SDSL-Solver achieves average speedups of $6.23\times$ and $7.77\times$ over PETSc under the Block Jacobi and BBD methods, respectively, and average speedups of $97.54\times$ and $5.85\times$ over the single-node PARDISO. Moreover, the solver remains numerically robust on ill-conditioned problems where direct methods fail to converge. These results establish that the synergy of carefully designed Krylov methods, adaptive preconditioning strategies, and architecture-aware parallel decomposition can effectively unlock the scalability of IPMs for industrial-scale optimization problems.

\section{Background}
\subsection{Interior Point Method}

Consider the standard-form linear programming problem:
\begin{equation}
  \min_{x}\; c^{T}x \quad \text{s.t.}\; Bx = b,\; x \geq 0, \label{eq:lp}
\end{equation}
where $B \in \mathbb{R}^{m \times n}$ is the constraint matrix, $c \in \mathbb{R}^{n}$ is the cost vector, and $b \in \mathbb{R}^{m}$ is the right-hand side.
Primal-dual interior point methods solve~\eqref{eq:lp} by applying Newton's method to a perturbed system of the Karush--Kuhn--Tucker (KKT) optimality conditions~\cite{wright1997primal}:
\begin{equation}
  \begin{pmatrix} B^{T}\lambda + s - c \\ Bx - b \\ XSe - \mu e \end{pmatrix} = 0,
  \label{eq:kkt}
\end{equation}
where $\lambda \in \mathbb{R}^{m}$ is the dual variable, $s \in \mathbb{R}^{n}$ is the slack variable, $X = \operatorname{diag}(x)$, $S = \operatorname{diag}(s)$, $e$ is the all-ones vector, and $\mu > 0$ is a barrier parameter that is driven toward zero as the algorithm progresses.

Linearizing~\eqref{eq:kkt} at each iteration yields the \emph{augmented system}:
\begin{equation}
  \underbrace{\begin{pmatrix} -D & B^{T} \\ B & 0 \end{pmatrix}}_{A}
  \begin{pmatrix} \Delta x \\ \Delta \lambda \end{pmatrix}
  =
  \begin{pmatrix} r_{1} \\ r_{2} \end{pmatrix},
  \label{eq:augmented}
\end{equation}
where $D = X^{-1}S$ is a positive diagonal matrix. Because $D$ is diagonal, the system~\eqref{eq:augmented} can be reduced via Schur complement elimination to the \emph{normal equation system} (also called the reduced system):
\begin{equation}
  \underbrace{(BD^{-1}B^{T})}_{S_{\mathrm{r}}} \,\Delta \lambda = r_{2} + BD^{-1}r_{1},
  \label{eq:schur}
\end{equation}
after which $\Delta x$ is recovered by back-substitution. The matrix $S_{\mathrm{r}}$ is symmetric positive definite and typically much smaller than $A$, making it the preferred target for linear solvers.

However, two practical difficulties arise during the IPMs iteration. First, when the constraint matrix $B$ contains dense columns, the Schur complement $S_{\mathrm{r}}$ can become nearly dense, resulting in excessive fill-in and memory consumption. In practice, dense columns in $B$ are relocated to form a bordered structure to mitigate this effect~\cite{gondzio2012interior}. Second, and more critically, as $\mu \to 0$, the diagonal entries of $D$ span an increasingly wide range: elements corresponding to variables approaching their bounds grow toward infinity while others shrink toward zero. This causes the condition number of $S_{\mathrm{r}}$ to deteriorate dramatically, often by many orders of magnitude across the IPMs iterations. The progressive ill-conditioning poses the central numerical challenge for the linear solver: early iterations may be well-conditioned and easy to solve, while later iterations demand high-precision solvers that are robust to extreme spectral properties.

Since the linear system solver accounts for over 70\% of the total IPMs computation time, the choice of solver and its parallel scalability directly determine the practical applicability of IPMs to large-scale optimization problems.

\subsection{Krylov Subspace Method}
Krylov subspace methods constitute a prominent class of iterative algorithms for solving large-scale linear systems $Ax=b$, where $A \in \mathbb{R}^{n \times n}$ is typically large and sparse. 
Representative Krylov methods, including CG~\cite{cghestenes1952methods}, GMRES~\cite{gmressaad1986gmres}, BiCGSTAB~\cite{bicgstabvan1992bi}, GCR($m$)~\cite{gcreisenstat1983variational}, and MINRES~\cite{minrespaige1975minres} are extensively employed for large sparse linear systems arising in structural analysis, partial differential equations, computational fluid dynamics, circuit simulation, and saddle-point problems. These methods rely primarily on matrix-vector products and vector operations, both of which are inherently amenable to parallelization.


For ill-conditioned or stiff problems, the convergence of unpreconditioned Krylov methods may be unacceptably slow or may fail entirely. To accelerate convergence, Krylov methods are typically combined with preconditioning techniques~\cite{scott2023algorithms}, which transform the original linear system into an equivalent system possessing more favorable spectral properties. The fundamental idea is to introduce a nonsingular matrix $M$, termed the preconditioner, such that $M \approx A$ and $M^{-1}$ is computationally inexpensive to apply. Three standard preconditioning forms exist: left preconditioning, right preconditioning, and two-sided (split) preconditioning, formulated as \eqref{eq:leftprecond}, \eqref{eq:rightprecond}, and \eqref{eq:twosideprecond}, respectively.
\begin{equation}
 M^{-1}Ax=M^{-1}b \label{eq:leftprecond}
\end{equation}
\begin{equation}
 AM^{-1}y=b, x=M^{-1}y \label{eq:rightprecond}
\end{equation}
\begin{equation}
 M_{1}^{-1}AM_{2}^{-1}y=M_{1}^{-1}b, \quad x=M_{2}^{-1}y \label{eq:twosideprecond}
\end{equation}
The choice of preconditioning form depends on the specific problem characteristics and the Krylov method employed. For GMRES, right preconditioning is generally preferred as it minimizes the norm of the true residual~\cite{saad2003iterative}. In this work, we adopt right preconditioning throughout. We employ factorization-type algorithms, such as complete or incomplete $LU$, $LDL^T$ factorization, as the preconditioner in SDSL-Solver. In the following discussion, we will uniformly use the $LU$ factorization for illustration.

\subsection{Block Jacobi Distributed Parallel Method}\label{sec:bj}

Block Jacobi preconditioning is a matrix preconditioning technique used to accelerate the convergence of iterative methods. It takes several diagonal blocks of the matrix as the preconditioner, while ignoring the elements outside these blocks. These blocks are independent of each other in factorization and solution phases, yielding natural parallelism. 
In distributed systems, the matrix partitioning method corresponding to Block Jacobi preconditioning is the row-wise partitioning method, i.e., different row blocks of the matrix are assigned to different processes, and the diagonal blocks are used as preconditioning submatrices. Therefore, we refer to this distributed parallel method as Block Jacobi distributed parallelization in this work. 

The Block Jacobi Distributed Parallel method is based on the principle of \emph{divide and conquer}, as shown on the left of the Fig~\ref{fig:xsolver_arch}, including five kinds of operations. 

\textbf{First, partition.} The global coefficient matrix $A$ is partitioned row-wise into $N$ sub-blocks, and each process receives a sub-block, together with the corresponding segments of the right-hand side vector $b$.

\textbf{Second, get precondition matrix.} The global preconditioner $M$ is defined as the block diagonal of $A$, i.e., the dark-colored portion within each row block in the figure. 

\textbf{Third, factorization.} Since $M$ is block diagonal, its inverse can be applied in parallel by independently solving the local systems on each process. In SDSL-Solver, each local block is preconditioned using ILU(0) or IC(0) factorization.

\textbf{Fourth, solve the precondition system.} For preconditioned systems, the preconditioner $M$ is factorized into the product of a lower triangular matrix $L$ and an upper triangular matrix $U$. The Krylov iteration requires a sparse triangular solve (SpTRSV), solving $Ly=b$ for $y$, and solving $Ux=y$ for $x$. As $M$ is a block diagonal matrix whose blocks are independent of one another, SpTRSV involves no communication.

\textbf{Finally, other operations.} Within the Krylov iteration, all operations are executed in a distributed manner:

\begin{itemize}

  \item \textbf{Sparse matrix-vector product (SpMV):} Each process computes the local product. To handle dependencies on vector entries owned by neighboring processes, \emph{ghost cell} data is exchanged via point-to-point MPI communication, forming a distributed SpMV operation.

  \item \textbf{Inner products:} Each process computes a local partial inner product, and a global result is obtained through \texttt{MPI\_Allreduce}, ensuring all processes hold identical scalar parameters (e.g., step lengths, residual norms).

  \item \textbf{Vector updates:} Operations such as $x \leftarrow x + \alpha p$ are performed entirely locally without communication.
\end{itemize}

The Block Jacobi distributes both computation and memory uniformly across processes, enabling the solution of problems that exceed single-node capacity. It is conceptually straightforward, amenable to implementation within the MPI programming model, and admits flexible substitution of local solvers. However, a fundamental limitation is that the preconditioner neglects the off-diagonal coupling between blocks. When the matrix $A$ is not diagonally dominant, or is relatively ill-conditioned, the discarded coupling becomes increasingly significant, potentially inflating the number of Krylov iterations required for convergence. For these matrices, SDSL-Solver provides the BBD described in the following subsection.

\subsection{Bordered Block Diagonal (BBD) Distributed Parallel Method}\label{sec:bbd}

The Bordered Block Diagonal (BBD) method, also referred to as the arrowhead or block staircase decomposition, is a distributed parallel strategy that preserves global coupling information through a shared interface system. In contrast to Block Jacobi, which discards inter-block connections, BBD explicitly retains a ``border'' that captures the coupling between sub-domains.

In the BBD parallel method, the coefficient matrix $A$ and right-hand side $b$ should be reordered via the Nested Dissection algorithm in ParMETIS~\cite{1997Parmetis}. $A$ is permuted in bordered block diagonal form, as shown on the right of the Fig~\ref{fig:xsolver_arch}, where the dark-colored parts are nonsingular diagonal block matrices representing each sub-domain's interior, the light-colored parts are the border coupling blocks describing the connection between sub-domains and the global interface, The purple block in the bottom‑right corner is the interface block representing the internal coupling of the global boundary.

The BBD method follows a \emph{divide-and-coordinate} strategy that proceeds in five stages:

\textbf{First, permutation.} The global coefficient matrix $A$ is permuted as a bottom-right arrow. The global preconditioning matrix $M$ is $A$.

\textbf{Second, partition.} Each diagonal block, along with its associated bordering blocks sharing the same row or same column, is assigned to a single process. The purple block at the bottom‑right corner resides on process 1.

\textbf{Third, factorization.} Each process factorizes its local submatrix, and can use either complete or incomplete factorization depending on the required solution accuracy. The factorization algorithm needs to use Schur complement factorization. Note that the purple block on process 1 is not involved in the computation in this stage. Then, all processes except process 1 send their local Schur complement parts to process 1, and process 1 accumulates these Schur matrices to obtain the overall Schur matrix. This process can be accomplished using \texttt{MPI\_Reduce}. Finally, process 1 performs a factorization of the final Schur matrix, yielding the bottom-right portion of the global matrix factorization. Thus, the factorization of the global preconditioner is finished.

\textbf{Fourth, solve the precondition system.} 
Since the factorization result of the bottom‑right purple matrix and the right‑hand side vector are both stored exclusively on process 1, when solving the global sparse lower‑triangular system, each process must send its resulting vector to process 1 after completing its local SpTRSV computation. Similarly, when solving the global sparse upper‑triangular system, process 1 first performs the SpTRSV computation on the purple matrix and then broadcasts the result to the other processes. Upon receiving the data, the other processes execute their local SpTRSV computations in parallel.

\textbf{Finally, other operations.} Other computations of the Krylov subspace iteration methods, such as SpMV, dot products, etc., are performed by exchanging necessary data through inter-process communication, based on the data dependencies.

Compared to the Block Jacobi method, BBD preserves all nonzero entries in the preconditioner, allowing it to capture the full coupling structure of the original matrix. This makes BBD particularly effective for general or ill-conditioned matrices where Block Jacobi's convergence degrades. However, when the interface size is large, the Schur complement solve on process 1 can become a bottleneck, since it is typically dense. In SDSL-Solver, sparse filtering can be applied to alleviate this cost when needed.

\section{SDSL-Solver}\label{sec:methods}

This section presents the four key techniques in SDSL-Solver: two distributed parallel methods (Block Jacobi and BBD) and two preconditioning enhancement strategies (numerics-based sparse filtering and diagonal correction). Together with the preconditioner reuse mechanism, these techniques form a comprehensive framework for efficiently solving the sparse linear systems in IPMs.

\subsection{Dynamic Parallel Switching Strategy}\label{sec:framework}

While the Block Jacobi method offers superior parallelism, its retention of only diagonal block information and discarding of off-diagonal coupling leads to a degradation in preconditioning quality, particularly for ill-conditioned matrices, often resulting in solver divergence. The BBD method, by contrast, decomposes the global matrix into independent sub-blocks while preserving inter-block adjacency relationships, thereby yielding higher preconditioning quality at the expense of additional global communication required to coordinate the interface system.

In view of these complementary trade-offs, SDSL-Solver dynamically selects between Block Jacobi and BBD based on the problem characteristics: Block Jacobi is preferred for diagonally dominant matrices (e.g., PageRank-type problems), while BBD is employed for general or ill-conditioned problems (e.g., LP and Instance-class problems) that demand high-fidelity preconditioning. Algorithm~\ref{alg:adaptive_selection} presents the dynamic selection strategy.

\begin{algorithm}
\caption{Dynamic Parallel Switching Strategy for SDSL-Solver}
\label{alg:adaptive_selection}
\begin{algorithmic}[1]
\REQUIRE Linear system $Ax = b$ at each interior-point iteration
\ENSURE Solution $x$

\STATE $\text{method} \leftarrow \text{BlockJacobi}$
\STATE $\text{switched} \leftarrow \text{false}$

\FOR{each interior-point iteration $k = 1, 2, \ldots$}
    \LOOP
        \STATE Solve $Ax = b$ using \text{method}
        \IF{solver succeeds}
            \STATE \textbf{break}
        \ELSE
            \IF{not switched}
                \STATE $\text{method} \leftarrow \text{BBD}$
                \STATE $\text{switched} \leftarrow \text{true}$
            \ELSE
                \STATE Report solver failure and terminate
            \ENDIF
        \ENDIF
    \ENDLOOP
    
    \IF{converged}
        \STATE \textbf{break}
    \ENDIF
\ENDFOR
\RETURN $x$
\end{algorithmic}
\end{algorithm}

The dynamic switching parallel strategy initially employs the Block Jacobi method at the onset of the IPMs computation. Upon detecting a solver failure, it automatically transitions to the BBD method, which remains fixed for all subsequent IPMs iterations.

SDSL-Solver supports multiple preconditioning options, including ILU(0), ILUT, and complete LU factorization. In accordance with the characteristics of each parallel method, ILU(0) is typically employed within the Block Jacobi framework, whereas complete LU factorization is adopted within the BBD method to ensure sufficient preconditioning quality for ill-conditioned systems.

\subsection{Numerics-Based Sparse Filtering Algorithm}\label{sec:filtering}

In the large-scale sparse matrices arising from IPMs, a substantial proportion of off-diagonal entries exhibit very small absolute values relative to the diagonal. For instance, in the L2CTA3D benchmark, over 90\% of the nonzero entries possess small absolute values that contribute minimally to the matrix's spectral properties. The numerics-based sparse filtering algorithm exploits this observation to construct sparser preconditioners that are significantly cheaper to factorize and apply, while retaining sufficient accuracy for effective preconditioning.

Given the coefficient matrix $A$, the filtering algorithm constructs a sparsified matrix $\tilde{A}$ by dropping entries whose absolute values are small relative to both their row and column diagonal elements. Specifically, an off-diagonal entry $a_{ij}$ ($i \neq j$) is dropped if it satisfies:
\begin{equation}
  |a_{ij}| < \tau \cdot |a_{ii}| \quad \text{and} \quad |a_{ij}| < \tau \cdot |a_{jj}|,
  \label{eq:filter}
\end{equation}
where $\tau > 0$ is a user-specified filtering threshold. All diagonal entries are unconditionally retained. The filtered matrix $\tilde{A}$ is then used as the preconditioner:
\begin{equation}
  P = \tilde{A}.
  \label{eq:filter_precond}
\end{equation}

The rationale behind this approach is twofold. First, entries satisfying~\eqref{eq:filter} have negligible influence on the dominant eigenvalues and eigenvectors of $A$, so $\tilde{A}$ preserves the essential spectral properties of $A$. Second, the reduction in nonzero entries dramatically decreases both the factorization time and the triangular solve time of the preconditioner. For matrices where 90\% or more of the entries are small, the filtered preconditioner can be orders of magnitude cheaper to compute.

Since the filtering introduces a discrepancy between $A$ and $\tilde{A}$, SDSL-Solver uses a complete Cholesky factorization on the filtered matrix $\tilde{A}$ rather than an incomplete factorization, thereby minimizing additional precision loss during the factorization phase. The complete solution procedure is shown in Algorithm~\ref{alg:filtered_preconditioner_build}:

\begin{algorithm}
\caption{Construction of the filtering-based preconditioner}
\label{alg:filtered_preconditioner_build}
\begin{algorithmic}[1]
\REQUIRE Coefficient matrix $A$, filtering threshold $\tau > 0$
\ENSURE Preconditioner $P = \tilde{L}\tilde{L}^T$
\FOR{each off-diagonal entry $a_{ij}$ ($i \neq j$)}
    \IF{$|a_{ij}| < \tau \cdot |a_{ii}|$ \textbf{and} $|a_{ij}| < \tau \cdot |a_{jj}|$}
        \STATE discard $a_{ij}$ (set to $0$)
    \ELSE
        \STATE retain $a_{ij}$
    \ENDIF
\ENDFOR
\STATE Keep all diagonal entries $a_{ii}$ unchanged
\STATE Compute complete Cholesky factorization: $\tilde{A} = \tilde{L} \tilde{L}^T$
\STATE Set preconditioner $P = \tilde{L} \tilde{L}^T$
\RETURN $P = \tilde{L} \tilde{L}^T$
\end{algorithmic}
\end{algorithm}

The filtering threshold $\tau$ controls the trade-off between preconditioner quality and computational cost. In our experiments, $\tau = 10^{-3}$ provides a good balance for the tested benchmarks, achieving substantial speedups without compromising IPMs convergence.

\subsection{Diagonal correction algorithm}\label{sec:diagcorr}

As the IPMs iterates toward optimality, the diagonal scaling matrix $D = X^{-1}S$ introduces increasingly extreme values, causing the condition number of the reduced system $S_{\mathrm{red}}$ to grow by many orders of magnitude. Even after applying MC64 reordering and scaling~\cite{2001On,hsl_mc64}, which permutes and scales the matrix to maximize diagonal dominance and normalize diagonal entries to unit absolute value, the condition number may remain too large for an accurate iterative solution.

The diagonal correction technique provides a complementary approach to further improve conditioning. The method operates on the preconditioner matrix $P$ (which may already have been processed by MC64) by adding a small positive constant $\delta$ to the absolute value of each diagonal entry:
\begin{equation}
  \hat{p}_{ii} = p_{ii} + \delta \cdot \operatorname{sign}(p_{ii}),
  \label{eq:diagcorr}
\end{equation}
where $\delta > 0$ is a prescribed correction value. This modification strengthens the diagonal dominance of the preconditioner without altering the original coefficient matrix $A$ used in the Krylov iteration. The corrected matrix $\hat{P}$ is then factorized and used as the preconditioner.

The diagonal correction is applied \emph{only to the preconditioner}, not to the coefficient matrix itself. This is a deliberate design choice: modifying the preconditioner affects only the convergence rate of the Krylov iteration, whereas modifying the coefficient matrix would alter the solution. Since the Krylov method still operates on the original system $Ax = b$, the converged solution retains the accuracy determined by the convergence tolerance, regardless of the perturbation in the preconditioner.

In practice, SDSL‑Solver combines MC64 reordering and scaling with diagonal correction, as shown in Algorithm~\ref{alg:mc64_diagcorr}. In step 1, the MC64 algorithm is applied to obtain a reordered and scaled matrix $A' = P_r A Q_c$, where $P_r$ and $Q_c$ are the row and column scaling matrices, and the permutation is embedded. In step 2, the preconditioner is constructed as $P = A'$. If the sparse filtering is also applied, $P = \tilde{A}'$. In step 3, diagonal correction~\eqref{eq:diagcorr} is applied to $P$ to obtain $\hat{P}$. The value of $\delta$ is chosen empirically, typically as $10^{-12}$. In step 4, $\hat{P}$ is factorized and used within the Krylov subspace method. Importantly, in the Krylov subspace method, the right-hand side of the preconditioned system must be properly permuted and scaled using $R_r$ and $R_c$.

\begin{algorithm}
\caption{Diagonal Correction Applied to Preconditioner}
\label{alg:mc64_diagcorr}
\begin{algorithmic}[1]
\REQUIRE Original coefficient matrix $A$, diagonal correction parameter $\delta > 0$ 
\ENSURE Preconditioner $\hat{P}$

\STATE \textbf{Step 1: MC64 reordering and scaling}
\STATE Apply MC64 algorithm to matrix \(A\)
\STATE Obtain $P_r$, $Q_c$from MC64
\STATE $A' = P_r A Q_c$

\STATE \textbf{Step 2: Preconditioner construction}
\IF {filter is applied}
    \STATE $\tilde{A}' = \operatorname{filter}(A')$
    \STATE Set $P = \tilde{A}'$ 
\ELSE
    \STATE Set $P = A'$ 
\ENDIF

\STATE \textbf{Step 3: Diagonal correction}
\FOR{each diagonal entry $p_{ii}$ in $P$}
    \STATE $\hat{p}_{ii} = p_{ii} + \delta \cdot \operatorname{sign}(p_{ii})$
\ENDFOR

\STATE \textbf{Step 4: Factorization}
\STATE $(\hat{L}, \hat{U}) = \operatorname{Factorize}(\hat{P})$


\RETURN $\hat{P} = \hat{L}\hat{U}$ 
\end{algorithmic}
\end{algorithm}

This combined strategy has proven particularly effective for the Instance-class benchmarks, where direct solvers (including MKL PARDISO with Cholesky) fail to converge due to extreme ill-conditioning, but SDSL-Solver with MC64 plus diagonal correction achieves stable IPMs convergence.

\subsection{Preconditioner Reuse}\label{sec:reuse}

In IPMs, the coefficient matrix $S_{\mathrm{red}}$ changes at every nonlinear iteration due to the updated diagonal scaling $D$. However, between consecutive iterations, the changes in $S_{\mathrm{r}}$ are often small, particularly during the middle iterations of the IPMs where the barrier parameter $\mu$ decreases gradually. This observation motivates a \emph{preconditioner reuse} strategy that amortizes the expensive preconditioner construction across multiple IPMs iterations.

The strategy works as follows. At IPMs iteration $k$, SDSL-Solver constructs and factorizes a preconditioner $P^{(k)}$ from the current matrix $S_{\mathrm{r}}^{(k)}$. At the next iteration $k+1$, rather than rebuilding the preconditioner from scratch, SDSL-Solver reuses the symbolic factorization from iteration $k$ and performs only the numerical factorization with the updated matrix values (or, in the most aggressive mode, reuses both the symbolic and numerical factorization entirely). This process continues for a configurable number of iterations or until the Krylov solver detects convergence degradation.

Preconditioner reuse is applicable to instances that employ direct LU factorization as the preconditioner and can be categorized into two cases.
First, for problems exhibiting poor convergence, the preconditioning matrix undergoes MC64 reordering and diagonal correction prior to direct LU factorization. In this setting, the reordering and symbolic analysis phases (ordering, fill-in prediction, and memory allocation) are performed once and amortized over subsequent IPMs iterations.
Second, for well-conditioned problems without the need for numerical filtering, one may periodically perform symbolic and numerical factorizations within the IPMs iterations, reuse each factorization for several steps, and repeat this process until convergence

For benchmarks including NetworkPlan\_6 and L1\_sixm250obs, preconditioner reuse yields $3.5\times$ and $1.35\times$ additional speedup over the already-optimized iterative solver, respectively. Notably, for NetworkPlan\_6, this strategy also improves convergence behavior: the slight perturbation introduced by a stale preconditioner can serve as implicit regularization, mitigating the tendency of the solver to be destabilized by the increasingly ill-conditioned late-stage systems.

\section{Experiment}\label{sec:experiment}

\subsection{Experimental setup}\label{sec:setup}

\textbf{Hardware platforms.} We evaluate SDSL-Solver on two cluster configurations to demonstrate cross-architecture portability:

\begin{itemize}
  \item \textbf{X86 cluster:} 4 nodes, each equipped with CPUs at 2.6\,GHz, $4 \times 13 = 52$ cores per node, and 730\,GB memory.
  \item \textbf{Kunpeng cluster:} 4 nodes, each equipped with Kunpeng-920 CPUs at 2.6\,GHz, $4 \times 32 = 128$ cores per node, and 521\,GB memory.
\end{itemize}

\textbf{IPM solver framework.} All experiments are conducted within a production interior point method solver. To evaluate SDSL-Solver under realistic end-to-end conditions, we replace its internal sparse-solve module with SDSL-Solver while keeping the remainder of the IPM pipeline (KKT system assembly, step-length computation, convergence checking, etc.) unchanged.

\textbf{Baseline solvers.} We compare SDSL-Solver against two established baselines:
\begin{itemize}
  \item \textbf{Intel MKL PARDISO}~\cite{schenk2004solving}: A shared-memory direct solver employing Cholesky or LU factorization with multi-threaded parallelism, serving as the single-node baseline.
  \item \textbf{PETSc}~\cite{balay2019petsc}: A widely adopted distributed iterative solver framework that partitions the matrix via a row-wise decomposition strategy and employs Block Jacobi preconditioning with ILU(0) local solvers, serving as the multi-node baseline.
\end{itemize}


\textbf{Benchmarks.} We use two categories of test problems, summarized in Table~\ref{tab:benchmarks}:
\begin{itemize}
  \item \textbf{Block Jacobi benchmarks} (diagonally dominant): PageRank\_1m, PageRank\_5m, com-youtube, cit-patents, L2CTA3D, thk\_48, and thk\_63. These are LP problems from PageRank benchmark \cite{lu2023cupdlp} and Hans Mittelmann benchmark \cite{hans_lp} with diagonally dominant reduced matrices.
  \item \textbf{BBD benchmarks} (general or ill-conditioned): These are LP problems whose reduced matrices are highly ill-conditioned and require precise preconditioning. The sources include LP formulations of network planning problems, LP relaxations of Security-Constrained Unit Commitment (SCUC) problems, MIPLIB 2017 benchmark, and Hans Mittelmann benchmark.
\end{itemize}

\textbf{Experimental design.} Our evaluation is organized into two complementary groups of experiments:
\begin{itemize}
  \item \textbf{Single-node experiments (Section~\ref{sec:single_solve}):} Conducted on a single X86 node, these experiments isolate the contribution of each optimization technique in SDSL-Solver from the effects of inter-node communication and distributed parallelism. The baseline is Intel MKL PARDISO.
  \item \textbf{Multi-node experiments (Section~\ref{sec:multi_solve}):} Conducted on up to 4 nodes of both the X86 and Kunpeng clusters, these experiments assess the distributed parallel efficiency of SDSL-Solver in both single-solve and end-to-end IPM settings. Here, SDSL-Solver is compared against PETSc and PARDISO, to demonstrate the advantages of SDSL-Solver at scale and across heterogeneous architectures.
\end{itemize}
In short, the single-node experiments serve as an ablation study of SDSL-Solver's algorithmic components, whereas the multi-node experiments quantify its real-world benefit when integrated into OptVerse as a drop-in replacement for sparse solver.

\begin{table}[ht]
  \caption{Summary of benchmark problems. $n$ is the dimension of the reduced matrix $S_{\mathrm{red}}$, and $\mathrm{nnz}_L$ is the number of nonzeros in its lower triangle.}
  \label{tab:benchmarks}
  \centering
  \resizebox{0.5\textwidth}{!}{
  \small
  \begin{tabular}{lllccc}
    \toprule
    Category & Problem & Abbr. & $n$ & $\mathrm{nnz}_L$ & Type \\
    \midrule
    \multirow{7}{*}{Block Jacobi}
    & PageRank\_1m  & pg1m & 1{,}000{,}000 & 62{,}999{,}999 & PageRank \\
    & PageRank\_5m  & pg5m & 5{,}000{,}000 & 314{,}999{,}999 & PageRank \\
    & com-youtube   & comy & 1{,}134{,}890 & 5{,}975{,}248 & PageRank \\
    & cit-patents   & citp & 3{,}774{,}768 & 20{,}258{,}734 & PageRank \\
    & L2CTA3D       & L2C & 80{,}224   & 2{,}380{,}640 & Hans Mittelmann \\
    & thk\_48       & th48 & 1{,}224{,}309 & 17{,}068{,}753 & Hans Mittelmann \\
    & thk\_63       & th63 & 1{,}891{,}095 & 26{,}397{,}063 & Hans Mittelmann \\
    \midrule
    \multirow{14}{*}{BBD}
    & NetworkPlan\_2  & net2 & 1{,}267{,}995 & 7{,}069{,}833 & LP \\
    & NetworkPlan\_3  & net3 & 1{,}298{,}478 & 7{,}348{,}226 & LP \\
    & NetworkPlan\_6  & net6 & 487{,}999 & 3{,}050{,}273 & LP \\
    & NetworkPlan\_7  & net7 & 1{,}066{,}725 & 7{,}141{,}981 & LP \\
    & NetworkPlan\_8  & net8 & 1{,}413{,}175 & 8{,}191{,}812 & LP \\
    & NetworkPlan\_9  & net9 & 4{,}582{,}382 & 25{,}748{,}800 & LP \\
    & NetworkPlan\_10  & net10 & 3{,}379{,}888 & 21{,}271{,}365 & LP \\
    & NetworkPlan\_11  & net11 & 7{,}613{,}064 & 42{,}183{,}414 & LP \\
    & NetworkPlan\_18  & net18 & 361{,}815 & 2{,}332{,}710 & LP \\
    & ScucRelax\_1  & scu1 & 46{,}430 &1{,}283{,}378  & SCUC \\
    & ScucRelax\_3  & scu3 & 81{,}575  & 3{,}271{,}595 & SCUC \\
    & ScucRelax\_6  & scu6 & 63{,}066 & 3{,}856{,}612 & SCUC \\
    & ScucRelax\_7  & scu7 & 82{,}268 & 3{,}085{,}073 & SCUC \\
    & BuildingEnergy & buie & 221{,}348 & 1{,}539{,}614 & MIPLIB 2017 \\
    & fome13  & fo13 & 3{,}898 & 39{,}306 & Hans Mittelmann \\ 
    & L1\_sixm250obs & l1si & 66{,}077 & 364{,}998 & Hans Mittelmann \\
    \bottomrule
  \end{tabular}
  }
\end{table}

\subsection{Single-node Performance}\label{sec:single_solve}
To isolate the contribution of each optimization technique from hardware-specific effects, all experiments in this subsection are conducted on a single X86 node. SDSL-Solver employs IGCR($m$) as the Krylov solver with a relative residual tolerance of $10^{-8}$.
 
\subsubsection{Numerics-based Sparse Filtering}

\begin{table}[htbp]
\centering
\caption{Per-step solution time statistics with sparse filtering applied to the L2CTA3D benchmark.}
\label{tab:fil_l2cta3d}
\centering
\small
\begin{tabular}{lccccc}
\toprule
    IPMs\_Step & $\tau$ & $\mathrm{nnz}_L$ & $\frac{\mathrm{nnz}_L}{\mathrm{nnz}_L^{0}}$ & IGCR\_Iter & time(s)\\
    \midrule
    0    & 0     & 24216405 & 100\%      & 1         & 110 \\
    1    & 0     & 24216405 & 100\%      & 1         & 60  \\
    2    & 0     & 24216405 & 100\%      & 1         & 63  \\
    3    & 1e-3  & 6223363  & 26\%   & 3         & 23  \\
    4    & 1e-3  & 5506971  & 23\%   & 3         & 22  \\
    5    & 1e-3  & 3775742  & 16\%   & 3         & 16  \\
    6    & 1e-3  & 1956206  & 8\%    & 4         & 14  \\
    7    & 1e-5  & 1795021  & 7\%    & 3         & 13  \\
    8    & 1e-5  & 1653438  & 7\%    & 3       & 10  \\
    9    & 1e-5  & 1112597  & 5\%    & 3         & 10  \\
    10   & 1e-6  & 1059397  & 5\%    & 3       & 10  \\
    11   & 1e-8  & 1400258  & 6\%    & 2       & 10  \\
    12   & 1e-8  & 980210   & 4\%    & 2         & 8   \\
    13   & 1e-8  & 857933   & 4\%    & 4       & 8   \\
    \bottomrule
\end{tabular}
\end{table}

For matrices whose entry magnitudes vary widely across IPM iterations, such as L2CTA3D, numerical filtering can dramatically reduce the nonzero count of the preconditioning matrix and hence the factorization cost. Table~\ref{tab:fil_l2cta3d} traces the filtering behavior over the 13 IPM iterations required for L2CTA3D to converge. As the optimization progresses, the disparity in entry magnitudes grows monotonically, enabling increasingly aggressive filtering thresholds to be applied starting from the early iterations. When the resulting preconditioner quality degrades below an acceptable level, the filter is temporarily relaxed and re-applied at a subsequent iteration with a refined threshold.

Compared with PARDISO LDL$^T$ without preprocessing (916\,s), SDSL-Solver with sparse filtering reduces the total IPM solution time to 377\,s, yielding a $2.43\times$ speedup.

\subsubsection{Diagonal Correction}

\begin{table}[htbp]
  \centering
  \caption{Comparison of SDSL-Solver (with $\delta=10^{-12}$ diagonal correction) vs.\ PARDISO LDL$^T$ for end-to-end IPMs convergence. "Numerical Issues Detected" is an error status indicating solver failure within the IPMs framework.}
  \label{tab:diag_correct}
  \centering
  \small
  \begin{tabular}{lcccc}
    \toprule
    & \multicolumn{2}{c}{\textbf{PARDISO LDL$^T$}} & \multicolumn{2}{c}{\textbf{SDSL-Solver}} \\
    \midrule
    Case & Iteration & Status & Iteration & Status \\
    net2 & 52 & Numerical Issues Detected & 57 & Optimal \\
    net3 & 27 & Numerical Issues Detected & 35 & Optimal \\
    net6 & 50 & Numerical Issues Detected & 60 & Optimal \\
    net7 & 69 & Numerical Issues Detected & 82 & Optimal \\
    net8 & 66 & Numerical Issues Detected & 73 & Optimal \\
    net9 & 81 & Numerical Issues Detected & 92 & Optimal \\
    net10 & 53 & Numerical Issues Detected & 52 & Optimal \\
    net11 & 61 & Numerical Issues Detected & 94 & Optimal \\
    net18 & 54 & Numerical Issues Detected & 62 & Optimal \\
    scu1 & 72 & Numerical Issues Detected & 35 & Optimal \\
    scu3 & 100 & Numerical Issues Detected & 35 & Optimal \\
    scu6 & 29 & Numerical Issues Detected & 50 & Optimal \\
    scu7 & 107 & Numerical Issues Detected & 66 & Optimal \\
    buie & 36 & Numerical Issues Detected & 56 & Optimal \\
    fo13 & 20 & Numerical Issues Detected & 35 & Optimal \\
    \bottomrule
  \end{tabular}
\end{table}

Diagonal correction lowers the effective condition number of the preconditioner, thereby accelerating iterative convergence and improving numerical robustness. As shown in Table~\ref{tab:diag_correct}, IPM terminated due to numerical issues for all 15 benchmarks when using PARDISO LDL$^T$, whereas SDSL-Solver with diagonal correction ($\delta = 10^{-12}$) successfully converges to the optimal solution in every case.

\subsubsection{Preconditioner Reuse}

We evaluate the preconditioner reuse strategy described in Section~\ref{sec:reuse}. Table~\ref{tab:reuse} compares the total IPM solution time with and without reuse on the X86 cluster. By amortizing the cost of symbolic analysis and numerical factorization across consecutive IPM iterations, the reuse strategy yields substantial speedups.

\begin{table}[htbp]
\centering
\caption{Effect of preconditioner reuse on total IPMs solution time (seconds) on the X86 cluster.}
\label{tab:reuse}
\centering
\small
\begin{tabular}{lccccc}
\toprule
    \multirow{2}{*}{Problem} & \multicolumn{2}{c}{Without Reuse} & \multicolumn{2}{c}{With Reuse} & \multirow{2}{*}{Speedup} \\
    \cmidrule(lr){2-3} \cmidrule(lr){4-5}
    & {IPMs\_Step}     & {Time}   & {IPMs\_Step}     & {Time}   & \\
    \midrule
    net6 & 60 & 441.32 & 60 &125.3 & 3.53 \\
    l1si & 53 &27.94 & 28 &20.70 & 1.35 \\
    \bottomrule
\end{tabular}
\end{table}

For NetworkPlan\_6, the IPM converges in 60 steps. MC64 reordering and symbolic factorization are performed once in the first step and reused from the 2nd through the 58th iteration; only the final two iterations require recomputation, while numerical factorization is updated at every step. This reuse strategy not only reduces the per-step computation time but also yields a $3.53\times$ overall speedup.

For L1\_sixm250obs, the matrix structure remains invariant throughout the solve, so symbolic factorization is performed only once. Numerical factorization is likewise computed at the first step and reused for the subsequent 10 steps, after which it is refreshed at every iteration. Notably, the slight perturbation introduced by the stale preconditioner acts as implicit regularization for the increasingly ill-conditioned late-stage systems, reducing the total IPM step count from 53 to 28 and delivering a $1.35\times$ speedup.

\subsection{Multi-node Performance}
\label{sec:multi_solve}

We evaluate distributed performance along two axes: \emph{single-solve} performance, which measures the wall-clock time of an individual linear system solve, and \emph{end-to-end} performance, which captures the total execution time from IPM initialization to termination. Single-solve experiments on the X86 platform compare SDSL-Solver against PETSc and PARDISO LDL$^T$; end-to-end experiments are conducted on both the X86 and Kunpeng platforms.

For each solver, we report the best wall-clock time across all tested process/thread configurations. Note that, unlike PETSc, which employs a pure MPI model, SDSL-Solver adopts a hybrid MPI+OpenMP execution model combining inter-node message passing with intra-node multi-threading.

Because the IPM computations for the NetworkPlan-class problems fail to converge under both PARDISO LDL$^T$ and PETSc, we restrict the evaluation on these benchmarks to single-solve performance at selected IPM steps and omit end-to-end comparisons.

SDSL-Solver pairs BiCGSTAB with the Block Jacobi framework for well-conditioned problems and IGCR($m$) with the BBD framework for ill-conditioned problems, both using a relative residual tolerance of $10^{-8}$.

\subsubsection{Single-solve Performance of Block Jacobi Method}

Table~\ref{tab:bj_x86} reports the single-solve speedup of SDSL-Solver over PETSc (4 nodes) and PARDISO LDL$^T$ (1 node) for the Block Jacobi benchmarks on the X86 cluster.
SDSL-Solver consistently outperforms PETSc across all benchmarks, primarily due to its sparse filtering optimization ($\tau = 10^{-3}$). The gains over PARDISO LDL$^T$ are even more pronounced: beyond the additional computational resources available to the distributed solver, PARDISO LDL$^T$ suffers from massive nonzero fill-in during factorization, which dominates its execution time.

On up to four compute nodes, SDSL-Solver achieves speedups of up to $53.27\times$ over PETSc, with a geometric mean of $6.23\times$. Relative to single-node PARDISO LDL$^T$, the speedups reach $198.25\times$ with a geometric mean of $97.54\times$. Notably, PARDISO LDL$^T$ fails entirely on PageRank\_1m, PageRank\_5m, com-youtube, and cit-patents because the fill-in volume exceeds the 32-bit integer indexing limit.

\begin{table}[t]
  \caption{Single-solve speedup of SDSL-Solver(Block Jacobi) over PETSc and PARDISO LDL$^T$ (abbreviated as PDS) up to 4 nodes of the X86 clusters. Step denotes the IPMs step, and Iter denotes the Bicgstab iteration. }
  \label{tab:bj_x86}
  \centering
  \resizebox{0.46\textwidth}{!}{
  \small
  \begin{tabular}{lccrcrrrr}
    \toprule
    \multirow{2}{*}{Problem} & \multirow{2}{*}{Step} & \multicolumn{2}{c}{SDSL\_Solver} & \multicolumn{2}{c}{PETSc} & PDS & \multicolumn{2}{c}{Speedup}\\ 
    \cmidrule(lr){3-4} \cmidrule(lr){5-6} \cmidrule(lr){7-7} \cmidrule(lr){8-9}
    & & Iter & Time & Iter & Time & Time & vs.PETSc & vs.PDS \\
    \midrule
    pg1m & 1 & 15 & 2.31 &  15 & 6.69 & -- & $2.90\times$ & -- \\
    pg5m & 1 & 15 & 13.63 &  16 & 44.86 & -- & $3.29\times$ & -- \\
    comy & 1 & 19 & 20.65 &  17 & 1099.86 & -- & $53.27\times$ & -- \\
    citp & 1 & 25 & 8.10 & 31 & 16.14 & -- & $2.35\times$ & --\\
    \midrule
    \multirow{3}{*}{L2C}
    & 1  &  7 & 0.41  & 7 & 0.85 & 80.49 & $2.09\times$ & $198.25\times$ \\
    & 5  &  15 & 0.51  & 15 & 1.05 & 162.87 & $2.06\times$ & $319.35\times$\\
    & 10 &  244 & 2.26  & 225 & 6.62 & 166.05 & $2.74\times$ & $73.47\times$\\
    \midrule
    \multirow{4}{*}{th63}
    & 1 & 877 & 8.38 & 888 & 15.89 & 66.55 & $1.90\times$ & $7.94\times$\\
    & 5  & 733 & 7.96 & 729 & 13.06 & 121.02 & $1.64\times$ & $15.20\times$\\
    & 10 & 812 & 8.77 & 749 & 13.06 & 119.30 & $1.49\times$ & $13.60\times$\\
    & 15 & 1401 & 14.14 & 1544 & 22.65 & 120.61 & $1.60\times$ & $8.53\times$ \\
    \midrule
    \multirow{3}{*}{th48}
    & 1 & 170 & 6.20 & 156 & 19.64 & 1221.75 & $3.17\times$ & $197.09\times$\\
    & 5 & 738 & 27.23 & 596 & 68.93 & 1209.67 & $2.53\times$ & $44.42\times$ \\
    \midrule
    Average &  & & & & & & $6.23\times$ & $97.54\times$\\
    \bottomrule
  \end{tabular}
  }
\end{table}

\subsubsection{Single-solve Performance of BBD Method}

\begin{table}[t]
  \caption{Single-solve speedup of SDSL-Solver(BBD) over PETSc and PARDISO LDL$^T$ (abbreviated as PDS) up to 4 nodes of the X86 clusters. Step denotes the IPMs step, and Iter denotes the Bicgstab iteration. }
  \label{tab:bbd_x86}
  \centering
  \resizebox{0.46\textwidth}{!}{
  \small
  \begin{tabular}{lccrcrrrr}
   \toprule
   \multirow{2}{*}{Problem} & \multirow{2}{*}{Step} & \multicolumn{2}{c}{SDSL\_Solver} & \multicolumn{2}{c}{PETSc} & PDS & \multicolumn{2}{c}{Speedup}\\ 
    \cmidrule(lr){3-4} \cmidrule(lr){5-6} \cmidrule(lr){7-7} \cmidrule(lr){8-9}
    & & Iter & Time & Iter & Time & Time & vs.PETSc & vs.PDS \\
    \midrule
    \multirow{2}{*}{net2}
    & 1 & 3 & 0.36 & 193 & 5.18 & 1.42 & $14.39\times$ & $3.94\times$ \\
    & 10 & 3 & 0.34 & 272 & 2.97 & 1.22 & $8.21\times$ & $3.59\times$ \\
    \midrule
    \multirow{2}{*}{net3}
    & 1 & 3 & 0.39 & 139 & 3.84 & 2.58 & $9.85\times$ & $6.62\times$ \\
    & 10 & 3 & 0.39 & 227 & 2.56 & 2.09 & $6.56\times$ & $5.36\times$ \\
    \midrule
    \multirow{2}{*}{net7}
    & 1 & 3 & 0.38 & 103 & 2.70 & 1.91 & $6.84\times$ & $5.03\times$ \\
    & 10 & 3 & 0.39 & 111 & 0.86 & 2.02 & $2.69\times$ & $5.18\times$ \\
    \midrule
    \multirow{2}{*}{net8}
    & 1 & 3 & 0.39 & 187 & 4.87 & 1.59 & $12.49\times$ & $4.08\times$ \\
    & 10 & 3 & 0.39 & 267 & 2.35 & 1.90 & $6.03\times$ & $4.87\times$ \\
    \midrule
    \multirow{2}{*}{net9}
    & 1 & 3 & 1.40 & 207 & 20.00 & 6.90 & $9.40\times$ & $4.93\times$ \\
    & 10 & 3 & 1.40 & 278 & 11.52 & 6.52 & $8.23\times$ & $4.66\times$ \\
    \midrule
    \multirow{2}{*}{net10}
    & 1 & 2 & 0.35 & 8 & 1.47 & 4.28 & $4.20\times$ & $12.23\times$ \\
    & 10 & 2 & 0.37 & 9 & 1.60 & 3.58 & $4.32\times$ & $9.68\times$ \\
    \midrule
    Average &  & & & & & & $7.77\times$ & $5.85\times$ \\
    \bottomrule
  \end{tabular}
  }
\end{table}

For the ill-conditioned NetworkPlan-class benchmarks, we evaluate SDSL-Solver with the BBD framework against PETSc and PARDISO LDL$^T$. Because the condition numbers of these matrices vary substantially across IPM iterations, we report results at multiple representative steps. Table~\ref{tab:bbd_x86} summarizes the single-solve speedups on the X86 cluster (4 nodes).

Against PETSc, SDSL-Solver achieves speedups ranging from $2.69\times$ to $14.39\times$, with a geometric mean of $7.77\times$. The primary contributor is the combination of diagonal correction and direct LU factorization in the preconditioner, which yields a higher-quality preconditioner and sharply reduces the IGCR iteration count.

Against PARDISO LDL$^T$, SDSL-Solver achieves a geometric mean speedup of $5.85\times$ (up to $12.23\times$). Although the direct LU factorization in SDSL-Solver is more expensive per process than PARDISO's LDL$^T$ factorization, the BBD parallel framework distributes the workload across multiple MPI processes with intra-process multi-threading, effectively offsetting the higher per-process cost.

We note that Table~\ref{tab:bbd_x86} includes only instances for which PETSc achieves a relative residual below $10^{-8}$; cases where PETSc fails to converge are excluded.


\subsubsection{End-to-end IPMs Performance}\label{sec:e2e}

We evaluate end-to-end IPM solution time on both the X86 and Kunpeng clusters with up to 4 nodes. Table~\ref{tab:e2e_results} reports the total wall-clock time for SDSL-Solver. PETSc is excluded from this comparison because it has not yet been integrated into the IPM framework for end-to-end evaluation.

For comy, pg5m, and th48, we adopt a two-phase filtering strategy: as the IPM progresses, the coefficient matrix entries evolve continuously, making a single threshold $\tau$ insufficient for the entire solve. Accordingly, $\tau$ is set to $10^{-3}$ for the first 24 steps and relaxed to $10^{-2}$ thereafter.

\begin{table}[t]
  \caption{End-to-end IPMs solve time (seconds) on the X86 cluster.}
  \label{tab:e2e_results}
  \centering
  \resizebox{0.45\textwidth}{!}{
  \small
  \begin{tabular}{lccrccr}
    \toprule
    \multirow{2}{*}{Problem} & \multicolumn{3}{c}{X86} & \multicolumn{3}{c}{kunpeng} \\
    \cmidrule(lr){2-4} \cmidrule(lr){5-7} 
    & $\tau$ & Steps & Time & $\tau$ & Steps & Time \\
    \midrule
    pg1m & $10^{-3}$ & 19 & 316.8 & $10^{-3}$ & 19 &523.5 \\
    citp & $10^{-3}$ & 40 &1{,}566.3 & $10^{-3}$ & 39 &2{,}210.4 \\
    comy & $10^{-3}/10^{-2}$ & 33 &2{,}351.0 & $10^{-3}/10^{-2}$& 33 &5{,}216.1 \\
    pg5m & $10^{-3}/10^{-2}$ & 31 &4{,}088.0 & $10^{-3}/10^{-2}$ & 31 & 9{,}989.0\\
    L2C & $10^{-3}$ & 12 &176.0 & $10^{-3}$ & 12 &382.6 \\
    th48 & $10^{-3}/10^{-2}$ & 31 &3{,}780.1 & -- & -- & -- \\
    th63 & $10^{-3}$ & 21 &1{,}589.0 & -- & -- & -- \\    
    \bottomrule
  \end{tabular}
  }
\end{table}

For the large-scale PageRank-type problems, PARDISO is inapplicable because the fill-in during factorization exceeds the 32-bit integer indexing limit. To put this in perspective, PARDISO requires 14{,}630\,s for a \emph{single} IPM iteration on the substantially smaller PageRank\_200k ($n{=}200{,}000$), whereas SDSL-Solver completes the \emph{entire} end-to-end solve for the $5\times$ larger PageRank\_1m in only 316.8\,s. This comparison highlights the fundamental scalability advantage of the distributed iterative approach over direct methods.

On the Hans Mittelmann benchmarks (L2CTA3D, thk\_48, thk\_63), SDSL-Solver achieves end-to-end speedups of $9.09\times$, $6.83\times$, and $1.32\times$ over single-node PARDISO on the X86 platform, consistent with the cumulative single-solve improvements across all IPM iterations. On the Kunpeng platform, thk\_48 and thk\_63 fail to converge to optimality. We attribute this to the progressive ill-conditioning of the coefficient matrix in later IPM iterations, compounded by the accumulation of floating-point rounding errors when computation is distributed across a larger number of cores, which collectively destabilize iterative convergence.




\subsection{Scalability Analysis}\label{sec:scalability}
\begin{figure}[t]
  \centering
  \includegraphics[width=0.85\columnwidth]{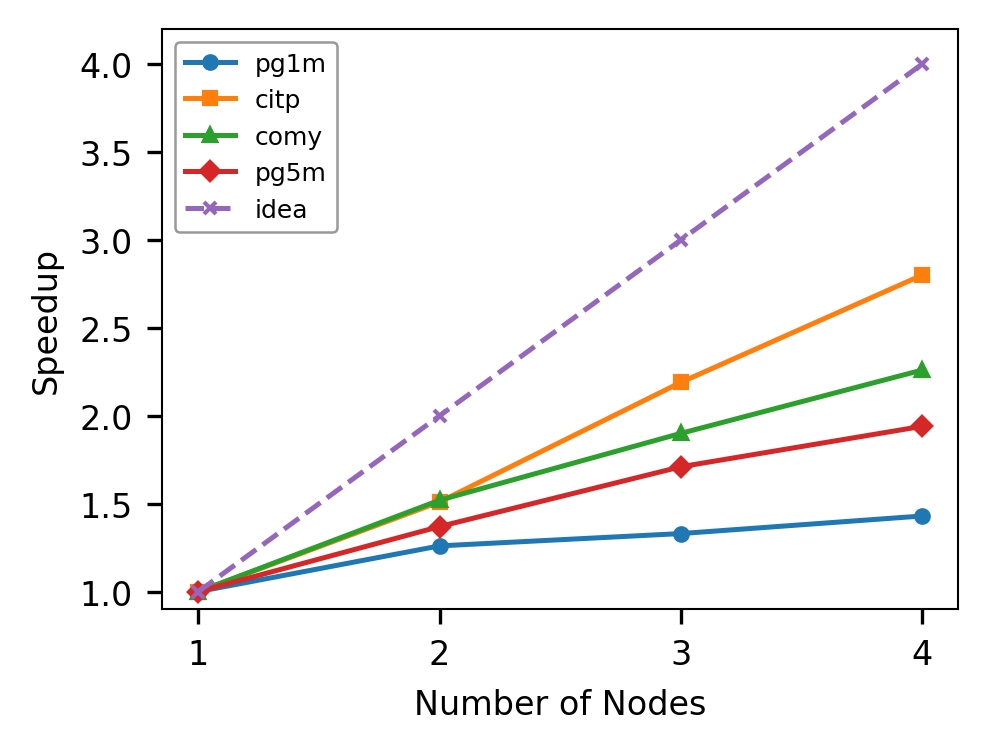}
  \caption{Strong scaling of SDSL-Solver on X86 cluster. The dashed line indicates ideal linear speedup.}
  \label{fig:scalability}
\end{figure}

Figure~\ref{fig:scalability} presents the strong scaling of SDSL-Solver on large-scale well-conditioned benchmarks from 1 to 4 nodes on the X86 cluster, with end-to-end IPM wall-clock time as the evaluation metric.
Because SDSL-Solver adopts a hybrid two-level MPI+OpenMP parallel model, the performance at a fixed core count can vary with the process/thread configuration. To ensure a fair comparison, for each node count, we sweep over all feasible process/thread combinations and report the configuration that minimizes end-to-end runtime.

As shown in Figure~\ref{fig:scalability}, the speedup (normalized to single-node runtime) grows consistently with the node count. The citp benchmark exhibits the best scaling, reaching $2.80\times$ on 4 nodes, while pagerank1m scales the worst, owing to its comparatively small problem size: as more resources are added, communication and task-distribution overheads cease to be amortized by sufficient computation. Moreover, non-solver components of the IPM pipeline are inherently difficult to parallelize. These two effects jointly erode parallel efficiency at higher node counts.


\section{Conclusion}\label{sec:conclusion}

We have presented SDSL-Solver, a scalable distributed sparse linear solver framework tailored to the large-scale, ill-conditioned systems that arise in interior point methods. SDSL-Solver couples the Krylov subspace solver with four complementary optimizations: (i) two distributed parallel strategies, Block Jacobi for diagonally dominant matrices and Bordered Block Diagonal for general or ill-conditioned matrices, that together cover the full spectrum of matrix difficulty in practical IPMs; (ii) a numerics-based sparse filtering scheme that exploits the magnitude distribution of IPM matrix entries to yield sparser yet effective preconditioners; (iii) A diagonal correction technique that improves the condition number of the preconditioning matrix for the extremely ill-conditioned systems encountered in late-stage IPM iterations; and (iv) a preconditioner reuse strategy that amortizes factorization costs across consecutive IPM iterations.

Extensive experimental results demonstrate that the multi-node SDSL-Solver attains a favorable speedup over both PETSc and the single-node PARDISO on matrices with millions of rows, while also substantially improving numerical robustness on benchmarks where direct solvers fail to converge. These results demonstrate that the Krylov methods with an adaptive preconditioner and dynamic parallel strategies may help  unlock the scalability of IPMs for industrial-scale optimization.

Future work will proceed along three directions: (i) Implement SDSL-Solver on GPU/NPU to obtain shorter computation time; (ii) adaptive threshold selection for sparse filtering guided by spectral estimates, thereby reducing manual parameter tuning; and (iii) extending SDSL-Solver to support second-order cone and semidefinite programming problems.


\bibliographystyle{ACM-Reference-Format}
\bibliography{sample-base}


\end{document}